\documentclass[conference]{IEEEtran}
\usepackage[breaklinks,plainpages=false]{hyperref}%
\usepackage{cite}
\usepackage{amsmath,amssymb,amsfonts}
\usepackage{graphicx}
\usepackage{textcomp}
\usepackage{xcolor}
\def\BibTeX{{\rm B\kern-.05em{\sc i\kern-.025em b}\kern-.08em
    T\kern-.1667em\lower.7ex\hbox{E}\kern-.125emX}}
    
\usepackage{cite}
\usepackage{tcolorbox}
\usepackage{xcolor,colortbl}

\definecolor{Gray}{gray}{0.85}
\definecolor{LightCyan}{rgb}{0.88,1,1}
\definecolor{DarkOrange}{rgb}{0.97, 0.72, 0.47} 
\usepackage{tikz}
\usetikzlibrary{fit,shapes.misc}

\newcommand\marktopleft[1]{%
    \tikz[overlay,remember picture] 
        \node (marker-#1-a) at (0,1ex) {};%
}
\newcommand\markbottomright[1]{%
    \tikz[overlay,remember picture] 
        \node (marker-#1-b) at (0,0) {};%
    \tikz[overlay,remember picture,thick,dashed,inner sep=3pt]
        \node[draw,rounded rectangle,fit=(marker-#1-a.center) (marker-#1-b.center)] {};%
}

\usepackage{caption}
\usepackage{subcaption}
\usepackage{tcolorbox}
\usepackage{booktabs}
\usepackage{multirow}
\usepackage{threeparttable}

\usepackage{balance}
\usepackage{algorithm}
\usepackage{algpseudocode}
\algnewcommand\algorithmicforeach{\textbf{for each}}
\algdef{S}[FOR]{ForEach}[1]{\algorithmicforeach\ #1\ \algorithmicdo}
\usepackage{hhline}

\begin{document}

\title{Machine Learning-based Automatic Annotation and Detection of COVID-19 Fake News}

\author{
\IEEEauthorblockN{Mohammad Majid Akhtar}
\IEEEauthorblockA{\textit{The University of New South Wales}\\
Sydney, Australia \\
majid.akhtar@unsw.edu.au}
\and
\IEEEauthorblockN{Bibhas Sharma}
\IEEEauthorblockA{\textit{The University of New South Wales}\\
Sydney, Australia \\
bibhas.sharma@home-in.com.au}
\and
\IEEEauthorblockN{Ishan Karunanayake}
\IEEEauthorblockA{\textit{The University of New South Wales}\\
Sydney, Australia \\
ishan.karunanayake@unsw.edu.au}
\and
\IEEEauthorblockN{Rahat Masood}
\IEEEauthorblockA{\textit{The University of New South Wales}\\
Sydney, Australia \\
rahat.masood@unsw.edu.au}
\and
\IEEEauthorblockN{Muhammad Ikram}
\IEEEauthorblockA{Macquarie University\\
Sydney, Australia \\
muhammad.ikram@mq.edu.au}
\and
\IEEEauthorblockN{Salil S. Kanhere}
\IEEEauthorblockA{\textit{The University of New South Wales}\\
Sydney, Australia \\
salil.kanhere@unsw.edu.au}
}

\maketitle

\begin{abstract}
COVID-19 impacted every part of the world, although the misinformation about the outbreak traveled faster than the virus. Misinformation spread through online social networks (OSN) often misled people from following correct medical practices. In particular, OSN bots have been a primary source of disseminating false information and initiating cyber propaganda. Existing work neglects the presence of bots that act as a catalyst in the spread and focuses on fake news detection in `articles shared in posts' rather than the post (textual) content. Most work on misinformation detection uses manually labeled datasets that are hard to scale for building their predictive models. In this research, we overcome this challenge of data scarcity by proposing an automated approach for labeling data using verified fact-checked statements on a Twitter dataset. In addition, we combine textual features with user-level features (such as followers count and friends count) and tweet-level features (such as number of mentions, hashtags and urls in a tweet) to act as additional indicators to detect misinformation. Moreover, we analyzed the presence of bots in tweets and show that bots change their behavior over time and are most active during the misinformation campaign. We collected 10.22 Million COVID-19 related tweets and used our annotation model to build an extensive and original ground truth dataset for classification purposes. We utilize various machine learning models to accurately detect misinformation and our best classification model achieves precision (82\%), recall (96\%), and false positive rate (3.58\%). Also, our bot analysis indicates that bots generated approximately 10\% of misinformation tweets. Our methodology results in substantial exposure of false information, thus improving the trustworthiness of information disseminated through social media platforms.
\end{abstract}


\section{Introduction}
As of August 2022, the entire world has been affected by COVID-19, leading to 581.8 million confirmed cases and 6.4 million deaths globally \cite{covid3}. 
Multiple lockdowns and workforce restrictions have impacted the entire world not only economically but also socially. This caused a paradigm shift for companies and educational institutes to adapt to a new ``work or study from home" operational model. A survey by Pew Research Center \cite{pewresearch} reveals that 90\% of American participants found the Internet personally important to them, and 40\% used it in a new way during the pandemic.

Social media is generally a rich source of information (or misinformation) for many. During the pandemic, many online users turned to it to get updated information. However, it resulted in an increase in misinformation (in this paper, we use the terms `misinformation', `false information' and `fake news' interchangeably) as well.
According to the World Health Organization (WHO), coronavirus misinformation is more contagious than the virus itself. Research shows that misinformation about the outbreak created a panic among the public \cite{depoux2020pandemic}. Examples of misinformation include blaming 5G technology as one of the reasons for the COVID-19 pandemic or accusing COVID-19 vaccines as a means for the government to control the population. These types of false campaigns confound emotional users into believing false remedies and neglecting the correct medical practices. Despite authorities and governments raising awareness and providing sufficient information to the public in a timely manner, there are still many myths around COVID-19 \cite{yang2020analysis}. 

In the recent past, researchers have used machine learning to detect misinformation or fake news in social media \cite{hatespeech, vicario2019polarization, antivaccination} and news articles \cite{Liar, wang2020weak, fakenewsnet}. However, most existing works only detect false information in OSN based on content with no emphasis on fake accounts that disseminate the false information in the initial phase of propagation spread \cite{shao2018spread}. Automated accounts, also known as ``bots'', actively spread misinformation, posing a severe threat to the genuine users of OSNs by hijacking public discussions and promoting their malicious goals~\cite{10}. Early identification of bots can prevent spread of misinformation \cite{arif2018acting}. Moreover, existing research has limitations in collecting labeled misinformation data efficiently as most works resort to manual annotation (or labelling) of data~\cite{hatespeech,antivaccination}. {\it Our work differs from studies as we automate the data labeling and identify misinformation in social media posts (text in posts) instead of fake news sources (articles shared in posts). Further, we analyze and detect the presence of OSN bots responsible for spreading misinformation}.

In essence, our work addresses the data scarcity problem in the existing literature by developing an automated machine learning based method for labeling data. In addition, we identify the presence of bots  posting misinformation tweets as they are a major part of false information spread. The main contributions of our work are as follows: 
\begin{itemize}
    \item  We propose an \textit{automatic annotation model} to annotate the textual content of tweets using supporting statements (verified fact-checked statements from government officials or experts). Essentially, our model uses a relevance matching (between tweets and supporting statements) machine learning model \cite{vo-lee-2020-facts} and a labeling algorithm to build an extensive ground truth dataset containing binary labels (\textit{fake} or \textit{real}) for information in the tweets.
    \item We used an ensemble model (combining multiple machine learning models) to detect fake news. We evaluated the applicability of various supervised learning classifiers such as K-Nearest Neighbour (KNN), Decision Tree (DT), Random Forest (RF), Naive Bayes (NB), Logistic Regression (LR) {for the} ensemble stack model. For every tweet, we used three types of features, i.e., tweet-level, user-level, and textual. To extract features, we used two different techniques, BERT Transformer model and Term Frequency-Inverted Document Frequency (TF-IDF). We found that the \textit{ensemble-based machine learning classifier} {consisting of Decision Tree, Random Forest and Logistic Regression} with TF-IDF performs best with high precision (82\%) and recall (96\%).
\item {We \textit{analyzed the presence of bots in our dataset} and found that bots generated approximately 10\% of misinformation tweets. Moreover, we compared the bot account's behavior from two periods (June-August 2021 and August 2022) based on bot score, a metric  to determine whether the account is a bot or not. We hypothesize that bots are more active during peak crises than at other times to maximize their goals. 5,315 unique accounts were identified as bots in June-August 2021 and the number is 3279 as of August 2022. This} depicts that bot behavior changes over time and is most active during misinformation campaigns (during crisis) compared to other times to maximize it's goals. We used \textit{Botometer Lite} \cite{botometerlite} to calculate the bot scores. 
\end{itemize}

The rest of the paper is structured as follows: Section~\ref{sec:datsetcollection} describes our data collection method and Section~\ref{sec:misinfo_annotationmodel} elaborates our proposed model for annotating misinformation. We leverage the annotated data to analyse and detect fake news and bots in Section~\ref{sec:fakenews_detection}, Section \ref{sec:experimental_results} and Section~\ref{sec:botdetection}, respectively. We present related work in Section~\ref{sec:relatedwork} and conclude our work in Section~\ref{sec:conclusion}. 

\section{Data Collection Methodology}
\label{sec:datsetcollection}
We begin by presenting our methodology for collecting COVID-19 related data from Twitter. 

{\bf Data Collection.} 
We collected tweets related to COVID-19 using a dataset from the Panacea Lab \cite{banda2021twitterdata}. Panacea Lab dataset contains roughly 730 million COVID-19 related tweet IDs. We use these tweet IDs to extract information, such as the tweet's textual content and metadata (by using Python libraries like Twarc \cite{twarc} and the Twitter API \cite{twitterAPI}). We used our institute's High-Performance Cluster (HPC) with multiple compute nodes, each having 1TB of memory, to conduct our data collection. Our data crawling process ran for two months, from June 2021 to August 2021. During this period, we collected tweets from 1st January 2020 to 21st June 2021 (533 days). Next, we filtered out the collected tweets based on a selection criteria, as explained below. 

{\bf Data Filtering.} 
Given that some countries may be exposed more to misinformation during the pandemic due to their higher Infection Rates (IRs), causing more panic among citizens \cite{depoux2020pandemic}. Therefore, we selected three geolocations, India, the United States of America (US), and the United Kingdom (UK), for their higher IRs and death counts. 
We also selected Australia as COVID-19 triggered the Australian government to impose a nationwide lockdown and border closure for a prolonged period. Next, we filtered out non-English tweets from the dataset of selected countries as this work is not language-agnostic. Our final dataset contained 10.22 Million tweets for further analysis.

{\bf Ethics Consideration.} 
We obtained ethics approval for data collection from our organization's ethics board. We do not intend to use, track, or de-anonymize users during our data collection, abiding the ethics guidelines in  \cite{rivers2014ethical}. Also, the data collected was not released publicly, and no personally identifiable information besides tweets and other metadata was collected.

\section{Misinformation Annotation Model}
\label{sec:misinfo_annotationmodel}
\begin{figure*}[!ht]
    \centering
    \includegraphics[width=12cm]{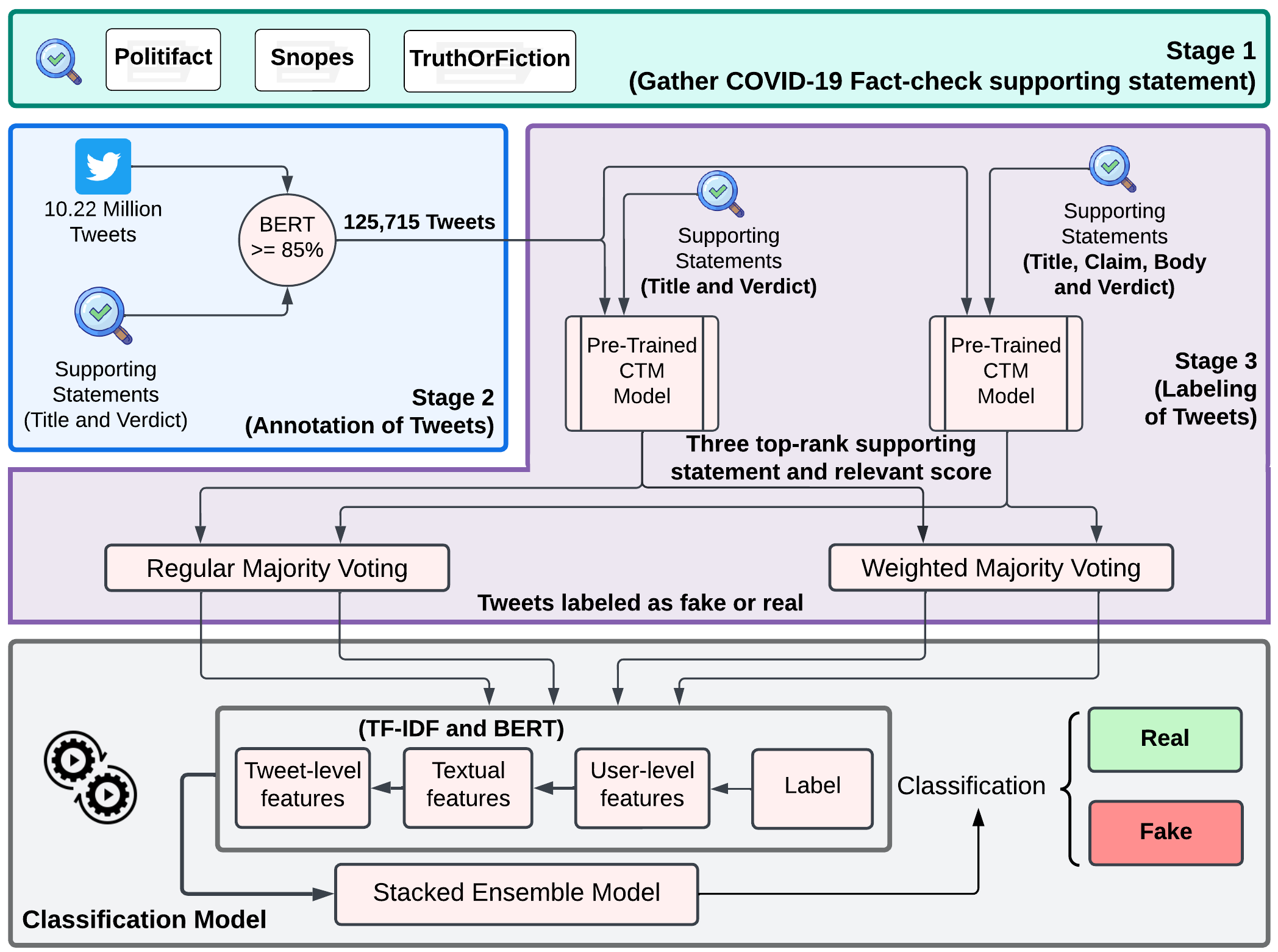}
    \caption{System Architecture for Fake News Detection. The architecture consists of different components as \textbf{Stage 1}: Data Gathering, \textbf{Stage 2}: Performing BERT-based comparison to annotate data, \textbf{Stage 3}: Labeling data using two algorithms, \textbf{Stage 4}: Passing data to Ensemble classifier for classification and detection of misinformation.
    }
    \label{fig:system_architecture}
\end{figure*}

This section provides an outline of the method used to annotate (or label) tweets in our dataset as \textit{fake} or \textit{real}. We employ end-to-end automation for annotation instead of manual expertise intervention. The model involves three different stages i.e., data gathering, filtering fake and real tweets (annotation), and labeling of tweets. Figure \ref{fig:system_architecture} shows the overall system architecture with different stages mentioned above. 

\textbf{Stage 1a. Gather COVID-19 Fact Checks:}  
Fact-checking websites such as \textit{Snopes}\footnote{https://www.snopes.com/}, \textit{Politifact}\footnote{https://www.politifact.com/} and \textit{TruthOrFiction}\footnote{https://www.truthorfiction.com/} examine trending news across social media and publish a verdict i.e., \textit{real} or \textit{fake}. They use verified and accredited sources, such as evaluations from journalists and experts \cite{fakenewsnet}. We collect this information from fact-checking websites~\cite{paka2021cross} and use it in our annotation model. In essence, we follow two data scraping approaches: \textit{without body text} and \textit{with body text}. In the first approach, we only extract the corresponding title and verdict, while in the second approach, the claim (statements tested for verifiability) and the body content (up to 350 words to add minimal auxiliary information) is extracted in addition to the title and the verdict. Our intuition behind this is to improve semantic matching using minimal auxiliary information and avoid over-matching raw similarities of common words like covid, vaccine, and virus. We also restricted scrapping up to 350 words to cater to memory requirements for algorithms used in sub-section \ref{sub-sec:CTM} (Stage 3). As a result, we collected 1,923  COVID-19 supporting statements for use in the next stage of the algorithm.

\textbf{Stage 1b. Grouping verdicts of COVID-19 Fact Checks:}
Many supporting statements--labels from the fact-checking websites--have different verdict classes, such as \textit{false, true, misattributed, pants-fire, etc.} As we are focusing on binary classification (i.e. \textit{fake} or \textit{real}), first, we need to group those different verdict classes into two. For example, we grouped together verdicts like pants-fire (from PolitiFact), misattributed (from Snopes), not true (from TruthOrFiction) and similar other untrue instances of information (full-flop, false, mostly-false) to the same category, \textit{fake}. We perform a similar step for the \textit{real} category to group verdicts like true, mostly-true, and correct attribution. Moreover, we discarded supporting statements with complex verdicts, such as \textit{unknown, unproven, satire, legend, research in progress}). This process resulted in 1,655 COVID-19 supporting statements with binary class labels, as shown in Table \ref{tb:supporting_statement}.

\begin{table}[h]
\centering
\caption{Number of COVID-19 Fact Checks supporting statements (SS) collected from fact-checking websites: Politifact, Snopes, and TruthOrFiction.}
\begin{tabular}{l|c|c|c|c}
\toprule
\toprule
&\textbf{Politifact} & \textbf{Snopes} & \textbf{TruthOrFiction} &\textbf{Total} \\ \hline
\shortstack{\textbf{\# Initial statements}}                      & 1,322              & 550    &     51  & 1,923                     \\ 
\shortstack{\textbf{\# After grouping}}                      & 1,218              & 386    &     51  & 1,655                     \\ 
\bottomrule
\bottomrule
\end{tabular}
\label{tb:supporting_statement}
\end{table}

\textbf{Stage 2. Filter real and fake tweets (Annotation):} In order to filter real tweets, we follow two approaches. First, we collect all tweets in the dataset posted by health organizations such as WHO, National Health Service (NHS), Governments' health portals or websites, etc. To get the tweet IDs of these organizations, we used the Twitter API. After collecting tweets, we annotate them as \textit{real} based on the assumption that the content posted by such health organizations is genuine. Next, we used the BERT transformer model \cite{devlin2018bert} to generate embeddings of the supporting statements (titles) and the tweets. We calculate the pairwise cosine distance between the supporting statement and the tweets with a defined threshold. To determine the threshold, we make use of a publicly available COVID-19 misinformation dataset \cite{patwa2021fighting} consisting of approximately 8,560 labeled tweets. In Table \ref{tb:cosineThres}, we observe that the threshold has an inverse relationship with the number of tweets annotated. For instance, with 0.70 as the threshold, 6,528 tweets are annotated. It means that when the cosine threshold is 0.70, only 6,528 tweets out of 8,560 have a similarity with our supporting document corpus. When the threshold is increased to 0.90, the number of tweets with a similarity to supporting documents reduce to 430.
 
\begin{table}[h]
\centering
\caption{Accuracy of the algorithm and number of annotated or labeled tweets across {\it cosine similarity} threshold values.}
\begin{tabular}{c|c|c}
\toprule
\toprule
{\bf Cosine Similarity} & {\bf Algorithm} &  \\ 
{\bf Threshold} & {\bf Accuracy} & {\bf \# Tweets annotated ($n$ = 8,560)} \\ 
\hline
{\bf 0.70}                      & 50.1\%              & {\bf 6,528}                                \\ 
0.80                      & 71.9\%              & 1,953                                \\ 
{0.85}                      & 91.7\%              & ~~~668                                 \\ 
{\bf 0.90}                      & 98.6\%              & {\bf~~~430}                                 \\ 
\bottomrule
\bottomrule
\end{tabular}
\label{tb:cosineThres}
\end{table}
 
In contrast, the \textit{threshold value has a linear relationship with accuracy}. With the same 0.70 threshold, we achieved 50\% accuracy, while it increases to 98\% with the threshold value of 0.90. Considering both these factors, 0.85 was deemed an appropriate threshold for pairwise cosine distance. We used this threshold of 0.85 to annotate the total corpus that yields 125,715 tweets resulting in 17,289 \textit{real} tweets and 108,426 \textit{fake} tweets. We named this annotated dataset as \textit{labeled data 0}. 

\textbf{Stage 3. {Labeling} Tweets:}\label{sub-sec:CTM} Opting for 0.85 as the threshold reduces the accuracy by 7\% when compared to a threshold of 0.90. In addition, we only consider the similarity with a single supporting statement in this method. Since we have three different fact-checking organization data, we can improve the accuracy of labeling if we can consider at least three supporting statements instead. In order to do that, we use two labeling algorithms to label the 125,715 tweets.

First, we used a pre-trained Multimodal Attention Network (MAN) \cite{vo-lee-2020-facts}, a relevance matching model for labeling tweets. A \textit{relevant score} is another metric used to measure the similarity between a tweet and supporting statements and determine if the supporting statement fact-checks the tweet. The MAN model is designed to focus on word interactions, semantic matching and contextual representations to give a \textit{relevant score} based on GloVe \cite{glove} and ELMo \cite{elmo} embeddings between a tweet and a supporting statement \cite{vo-lee-2020-facts}. The MAN model requires `$n$' initial candidate lists of supporting statements per tweet retrieved by BM25 \cite{bm25}. As we deal with textual data in this work, a MAN-variant Contextual Text Matching (CTM) model is used. The CTM model \cite{vo-lee-2020-facts} is pre-trained on 467 Politifact fact-check articles as the majority of supporting statements are from Politifact. 

Next, we used the pre-trained CTM model for testing. When a tweet ($t$) is tested using the CTM model, it gives three top-rank statements ($s$) from the supporting statements derived from BM25 (for $n$=50) that fact-checks the tweet with a relevant score $f(t,s)$. The higher the relevant score, higher the probability that the supporting statement is fact-checking the tweet. Furthermore, we removed six tweets from the total set of 125,715 tweets as a relevant score could not be generated for them. We tested our remaining (125,709) tweets with both forms of scrapped data (supporting statements) i.e. \textit{without body text} and \textit{with body text} using the CTM model and achieved three top-rank supporting statements for each tweet with their relevant score. We refer to the outcome of this process as the $tweet_{result}$.

Next, we need a method to determine the final label for the tweet using its $tweet_{result}$. We explore two different approaches to do that. First, a majority vote is calculated for the three top-rank statements in $tweet_{result}$ using only the verdict of each statement. For example, if two of the three statement's verdict in $tweet_{result}$ is fake, then the tweet is fake. We call this method as \textbf{regular majority voting}. In this method, both cosine similarity and relevant score is not taken into account and all three votes are equally weighted.  
{Second, we introduce weighted majority voting to give importance (weights) to supporting statements based on their cosine similarity and relevant scores in the $tweet_{result}$. For example, suppose rank 2 and 3 supporting statement's verdict is fake with a low relevant score and cosine similarity, but rank 1 verdict is {\it real} with a high relevant score and cosine similarity. In that case, the tweet is labeled as {\it real}. To calculate the weight of each vote, we need to use relevant score and cosine similarity between the tweet and each supporting statement.}

{Next, we iterate through three top-ranked supporting statements for each tweet and calculate two scores i.e., $score_{fake}$ and $score_{real}$ using the supporting statement's verdict, cosine similarity, and relevant score. For instance, if the first statement's verdicts is \textit{`fake'}, then we increase the value of $score_{fake}$ using the relevant score and cosine similarity (cosine similarity of the supporting statement is divided by the sum of cosine similarities of the top 3 statements. This number is then multiplied by the relevant score of that supporting statement and added to $score_{fake}$ (or $score_{real}$)).} Finally, we compare both $score_{fake}$ and $score_{real}$ to label the respective tweet. The algorithm for \textbf{weighted majority voting} is described in Algorithm \ref{alg:weighted_majority_voting}. Therefore, we assign a label to tweets using regular and weighted majority voting with scrapped supporting statements (\textit{without body text and with body text}). This results in four variety of labeled data (1-4) as shown in Table \ref{tab:train_test_set} to be used later in Section \ref{sec:evaluation}.

\begin{algorithm}[h]
\footnotesize 
\caption{\small Algorithm for Weighted Majority Voting}\label{alg:weighted_majority_voting}
\begin{algorithmic}[1]
\ForEach {tweet in $tweet_{result}$}
    \State $score\_fake \gets 0$
    \State $score\_real \gets 0$
    \State $cosine\_similarity\_sum \gets 0$
    \State $cosine\_sim = [\ ]$
    \State $tweet\_embedding \gets bert.encode(tweet)$
    \For {i in range(0,3)} \Comment{for each supporting\_statement (ss)}
    \State $ss\_embedding \gets bert.encode(ss[i])$
    \State $ sim = cosine\_similarity(tweet\_embedding, ss\_embedding)$
    \State $cosine\_similarity\_sum = cosine\_similarity\_sum + sim$
    \State $cosine\_sim.append(sim)$
    
    \EndFor
    
    \For {i in range(0,3)} \Comment{for each supporting\_statement}
    \State $verdict \gets verdict\_of\_supporting\_statement\_at\_rank\_i$
    \State $relevant\_score \gets relevant\ score\ of s\_statement\ at\ rank\ i$ 
    \If{verdict == `fake'} 
        \State $score\_fake \gets score\_fake + ((cosine\_sim[i]/cosine\_\newline similarity\_sum)*relevant\_score)$
    \ElsIf {verdict == `real'} 
        \State $score\_real \gets score\_real + ((cosine\_sim[i]/cosine\_\newline similarity\_sum)*relevant\_score)$
    \EndIf
    \EndFor
    \If{score\_fake $>=$ score\_real} 
        \State tweet in $tweet_{result} \gets fake$
    \Else
        \State tweet in $tweet_{result} \gets real$
    \EndIf
\EndFor
\end{algorithmic}
\end{algorithm}

\section{Proposed Fake News Detection Methodology}
\label{sec:fakenews_detection}
In this section, we describe our feature extraction process and the model architecture for fake news detection.  

\subsection{Feature Extraction}

We extract three different types of features for each tweet. These include tweet-level features, user-level features, and textual features. For tweet-level features, we consider a range of different attributes including - \textit{number of user mentions, number of hashtags, number of URLs, number of favourites, number of retweets, number of media, is reply (0,1), number of special characters} and \textit{tweet length}. For user-level features, we consider attributes such as \textit{is verified user (0,1), number of followers, number of friends, number of favourites} and \textit{number of statuses}. We performed an empirical analysis for the extraction of textual features with three vectorizers such as Bag of Words, TF-IDF and BERT. TF-IDF and BERT outperformed Bag of Words. Therefore, we applied TF-IDF and the BERT transformer model to extract features from the tweets. 

\subsection{Stack Ensemble Model Architecture} \label{sec:modelArchitecture}

\begin{figure}[!ht]
    \centering
    \includegraphics[width=\linewidth]{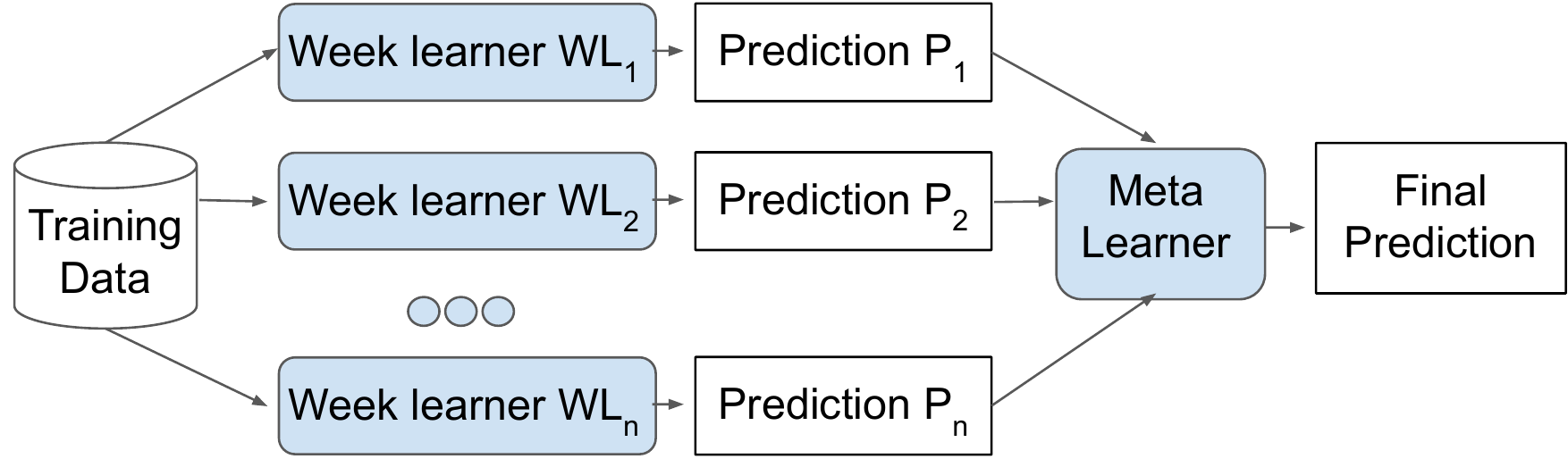}
    \caption{Overview of the proposed Stacked Ensemble model for detecting Fake News. }
    \label{fig:stack_model}
\end{figure}

Our proposed model architecture is an ensemble learning approach whereby multiple single machine-learning models (i.e. weak learners) are trained and a final meta-learner puts together the predictions of each weak learner to produce a final prediction as shown in Figure~\ref{fig:stack_model}. This increases the accuracy and enhances the model's ability to generalize a trend from the training samples.

For the weak-learners, we investigated multiple machine-learning models, including K-Nearest-Neighbour (KNN) \cite{KNN}, Decision Tree \cite{DecisionTree}, Random Forest \cite{RandomForest} and Naive Bayes \cite{naivebayes}, of which the high performing models are selected for our ensemble model. Similarly, we also compared the performance of the models (e.g. Logistic Regression, KNN, Decision Tree, and Random Forest) to select the meta-learner. Finally, we compared the performance of different combinations of weak-learners to fine-tune the overall model. We conducted this performance test using tweets and labels from stage 2 (\textit{labeled data 0}). Our classification model is shown in Figure \ref{fig:system_architecture}. 

\section{Experimental results}
\label{sec:experimental_results}
This section presents our results on the COVID-19 misinformation classification and detection task with our proposed model. 

\subsection{Performance comparison of Single Weak-learners}
\label{sec:single_weak_learners}

The first step is to determine the appropriate weak-learners for the proposed architecture of the ensemble classifier illustrated in section \ref{sec:modelArchitecture}. For that purpose, we selected multiple machine learning classifiers and individually compared the performances of each of those models to train them with the (\textit{labeled data 0}) dataset. We use precision and recall as the metrics to evaluate the classifiers. As the dataset is imbalanced, accuracy is not deemed as an ideal performance metric in this scenario. Table~\ref{tb:performWL} shows the performance of the four weak-learners, including KNN, Decision Tree, Random Forest, and Naive Bayes. We can see that Random Forest outperforms all the other models with a {\it precision} score of 92\% and {\it recall} of 92\%. Since Random Forest avoids overfitting using multiple decision trees and gives precise results. Moreover, Naive Bayes performed poorly; hence, it was not considered for the rest of the experiments.

\begin{table}[!ht]
\centering
\caption{Performance comparison of single weak-learners.}
\label{tb:performWL}
\tabcolsep=0.20cm
\begin{tabular}{l|c|c|c}
\toprule
\hline
{\textbf{Weak-learner}} & {\textbf{Weighted Precision}} & {\textbf{Weighted Recall}} & {\textbf{F1-score}}  \\ \hline
KNN                                          & 82\%                                               & 86\% & 83\%                                           \\ 
Decision Tree                                & 87\%                                               & 89\% &87\%                                           \\ 
\textbf{Random Forest}                       & \textbf{92\%}                                      & \textbf{92\%}   & \textbf{91\%}                              \\ 
Naive Bayes                                  & 77\%                                               & 84\% & 80\%                                         \\ 
\hline
\bottomrule
\end{tabular}
\end{table}

\subsection{Determining the Appropriate Meta-learner}

Next, we trained four different machine learning classifiers as the meta-learner and compared their performance to determine the best meta-learner based on the overall performance of the proposed ensemble learning model. Specifically, we selected Logistic Regression, KNN, Random Forest, and Decision Tree as the meta-learner. As weak-learners, we selected KNN, Decision Tree and Random Forest at this experiment stage. The performance of candidate meta-learners are illustrated on Table \ref{tb:performML}. It is apparent that Logistic Regression is an appropriate selection of meta-learner as it outperforms all the other models. Since Logistic Regression combines probabilities along with classification and is less prone to over-fitting, hence, Logistic Regression was chosen as the meta-learner for the ultimate model.

\begin{table}[h]
\centering
\caption{Performance comparison of meta-learners.}
\label{tb:performML}
\scalebox{0.9} {
\begin{tabular}{l|c|c|c}
\toprule
\hline
{\textbf{Meta-learner}} & {\textbf{Weighted Precision}} & {\textbf{Weighted Recall}} & {\textbf{F1-score}} \\ \hline
\textbf{Logistic Regression}                 & \textbf{92\%}                                      & \textbf{93\%}                                   & \textbf{92\%}                             \\ 
KNN                                          & 91\%                                               & 92\%                                            & 91\%                                      \\ 
Decision Tree                                & 92\%                                               & 92\%                                            & 92\%                                      \\ 
Random Forest                                & 92\%                                               & 92\%                                            & 92\%                                      \\ 
\hline
\bottomrule
\end{tabular}}
\end{table}

\subsection{Determining the Appropriate Weak-learner combination.}

\begin{figure}[!ht]
    \centering
    \includegraphics[width=\linewidth]{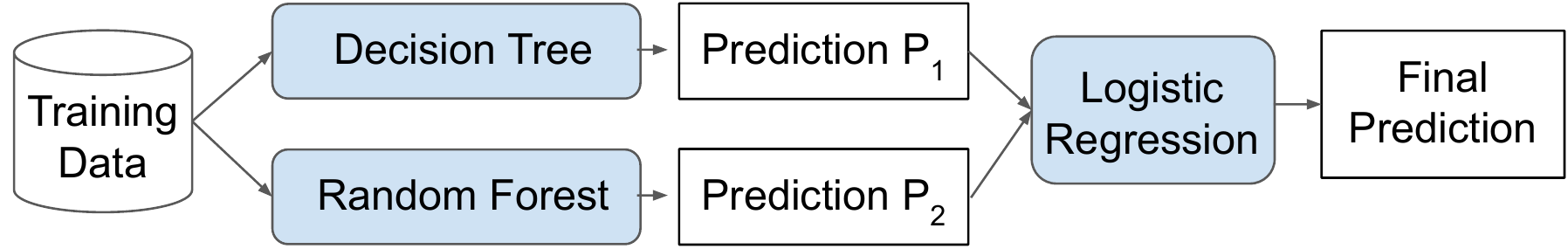}
    \caption{Overview of our methodology for detecting Fake News. The detection pipeline consists of ensemble of machine learning models (Decision Tree, Random Forest, and Logistic Regression) to predict fake news in the dataset.}
    \label{fig:ultimateModel}
\end{figure}

Herein, we selected combinations of three machine learning models and compared their performance to determine the appropriate combination of weak-learners. Table \ref{tb:performCombinationWL} shows the performance comparison of the combination of weak-learners with the selected meta-learner, i.e., Logistic Regression. It is apparent that all the combinations apart from KNN+Decision Tree have performed equally. It correlates with the fact that KNN involves distance calculation with each existing point, degrading the performance of a model with a large dataset. For the sake of simplicity of the model and faster training and testing , we selected Decision Tree and Random Forest as the weak-learners for the final model. Figure \ref{fig:ultimateModel} illustrates our final model for detecting misinformation.

\begin{table}[h]
\centering
\caption{Performance comparison of combination of weak-learners. Here \textit{WP} and \textit{WR} represent Weighted Precision and Weighted Recall. }
\label{tb:performCombinationWL}
\scalebox{1.0} {
\begin{tabular}{l|c|c|c}
\toprule
\hline
{\textbf{Weak-learners}} & {\textbf{WP}} & {\textbf{WR}} & {\textbf{F1- score}} \\ 
\hline
KNN + Decision Tree                           & 87\%                                               & 89\%                                            & 87\%                                      \\ 
KNN + Random Forest                           & 92\%                                               & 93\%                                            & 92\%                                      \\ 
\textbf{Decision Tree + Random Forest}        & \textbf{92\%}                                      & \textbf{93\%}                                   & \textbf{92\%}                             \\ 
All Models                                    & 92\%                                               & 93\%                                            & 92\%                                      \\ 
\hline
\bottomrule
\end{tabular}}
\end{table}

\subsection{Performance evaluation of our Model}
\label{sec:evaluation}
For the evaluation, we use the \textit{labeled data 1-4} datasets. Before conducting the experiments, we split each dataset into 80-20, where 80\% and 20\% of the sample were used for training and testing, respectively. The train and test set is shown in Table \ref{tab:train_test_set}. 

\begin{table*}[h]
  \centering
   \caption{Train and Test Set. We first named our four datasets (as column \textit{Named as}). We split our four labeled datasets in an 80:20 split approach for the different classifier models.}
\label{tab:train_test_set}
  \begin{tabular}{c c|c|c|c|c|c|c}
   \toprule
   \hline
   \multicolumn{2}{c|}{\multirow{2}{*}{\textbf{Labeled Data Description}}}&\multirow{2}{*}{\textbf{Named as}}&\multirow{2}{*}{\textbf{Total Tweets}}&\multicolumn{2}{c}{\textbf{Train (80\%)}}&\multicolumn{2}{c}{\textbf{Test (20\%)}}\\\hhline{~~~~----}
  & & & &\textbf{Fake} & \textbf{Real} &\textbf{Fake} &\textbf{Real}\\
   \hline
   \multicolumn{2}{c|}{\textbf{cosine distance $>=$ 0.85}}& Labeled Data 0 & 125,715 & 86,738 & 13,834 & 21,688 & 3455
   \\ \hline
   \multirow{2}{*}{\textbf{\shortstack{Without\\ body text}}}&\shortstack{Regular Majority}&Labeled Data 1& 125,709&74,638 &25,929 &18,710 &6,432\\
   \hhline{~-------}
   &\shortstack{Weighted Majority}&Labeled Data 2&125,709&74,008 &26,559 &18,611 &6,531\\
   \hline
   \multirow{2}{*}{\textbf{\shortstack{With\\ body text}}}&Regular Majority&Labeled Data 3&125,709&77,200 &23,367 &19,261 &5,881\\
   \hhline{~-------}
   &\shortstack{Weighted Majority}&Labeled Data 4&125,709&75,697 &24,870 &18,903 &6,239\\
   \hline
   \bottomrule
  \end{tabular}
\end{table*}

Table \ref{tab:results} shows the performance comparison of our stack ensemble model evaluated on our four datasets (described in Table \ref{tab:train_test_set}). As already discussed in Section \ref{sec:single_weak_learners}, we have used \textit{precision}, \textit{recall} and \textit{F1-score} as evaluation metrics. The values in Table \ref{tab:results} represent precision, recall and F1-score for \textit{fake} class, as our primary objective is to classify fake tweets as fake. We are also interested in a higher recall value, which means that few fake tweets are misclassified as real information. In the last column, we present the false positive rate (FPR) to show the ratio of fake news wrongly classified as real news. Overall, we obtain the best precision, recall and $F_1$ score values for \textit{labeled data 3} when tweet features were extracted using TF-IDF. The recall value is as high as 96\% in that case. Both TF-IDF and BERT almost perform similarly in all cases. However, for labeled data 2,3 and 4, BERT performed slightly lower than TF-IDF in recall and F1 score in Random Forest. It also achieved lower precision than TF-IDF in labeled data 3 in Stacked Model. In contrast, TF-IDF performed better or similar to BERT in all four labeled data. Since our work is event-specific, i.e., specific to Covid-related, it leads to fewer contextual differences that BERT is trying to learn. However, providing more body words than 350 might increase BERT performance more than TF-IDF. 
{We also compared the execution time of TF-IDF and BERT for labeled data 3. TF-IDF took 274 seconds while BERT took 663 seconds. Therefore, TF-IDF was approximately 2.5 times faster than BERT. Furthermore, data labeled using weighted majority voting (\textit{labeled data 2 and 4}) has slightly low or equal precision as compared to regular majority voting (\textit{labeled data 1 and 3 respectively}) in all models}. This is because the weighted majority method labels fewer fake instances than the regular majority one thus reduces the number of indicators for detecting fake news. Table \ref{tab:train_test_set} highlights results of our proposed approach.

\begin{table*}[h]
  \centering
   \caption{Performance of our proposed {\it model}. Here $P$, $R$, $F_1$, and $FPR$ represent precision, recall, $F_1$ score, and false postive rate. We compare and measure the performance of our model across different datasets:  \textbf{Regular Majority Voting}: Majority out of three supporting statement verdict's; 
       \textbf{Weighted Majority Voting}: Voting based on rank and relevant score of supporting statement;
      \textbf{Without body text}: Supporting statement's title and verdict; 
       \textbf{With body text}: Supporting statement's title, claim, body (up to 350 words) and verdict.}
\label{tab:results}
  \begin{tabular}{c c c|c|c|c|c|c|c|c|c|c|c}
   \toprule
   \hline
   \multicolumn{3}{c|}{\multirow{2}{*}{\textbf{Labeled Data Description}}}&\multicolumn{3}{c|}{\textbf{Decision Tree}}&\multicolumn{3}{c|}{\textbf{Random Forest}}&\multicolumn{4}{c}{\textbf{Stacked Model}}\\\hhline{~~~----------}
  & & &\textbf{P} & \textbf{R} &\textbf{F$_1$} &\textbf{P} & \textbf{R} &\textbf{F$_1$} &\textbf{P} & \textbf{R} &\textbf{F$_1$}&\textbf{FPR} \\
   \hline
   \multirow{4}{*}{\textbf{\shortstack{Without\\ body\\text}}}&\multirow{2}{*}{\shortstack{\textbf{Regular Majority}\\(\textit{Labeled Data 1})}}&\textbf{TF-IDF}  &75\% &100\% &86\% &77\% &98\% &87\% &77\% &98\% &86\%&2.03\%\\
   \hhline{~~-----------}
   &&BERT &76\% &99\% &86\% &77\% &96\% &86\% &77\% &98\% &86\%&1.86\%\\
   \hhline{~------------}
   &\multirow{2}{*}{\shortstack{Weighted Majority\\(\textit{Labeled Data 2})}}&TF-IDF &75\% &100\% &85\% &77\% &98\% &86\% &77\% &97\% &86\%&3.22\%\\
   \hhline{~~-----------}
   &&BERT &75\% &99\% &85\% &77\% &95\% &85\% &77\% &98\% &86\%&1.98\%\\
   \hline
   \multirow{4}{*}{\textbf{\shortstack{With\\ body\\text}}}&\multirow{2}{*}{\shortstack{Regular Majority\\(\textit{Labeled Data 3})}}&TF-IDF &\textbf{80\%} &\textbf{98\%} &\textbf{88\%} &\textbf{81\%} &\textbf{98\%} &\textbf{89\%} &\cellcolor{blue!15}\marktopleft{c1}\textbf{82\%} &\cellcolor{blue!15}\textbf{96\%} &\cellcolor{blue!15}\textbf{89\%}&\cellcolor{blue!15}\textbf{3.58\%}\markbottomright{c1}\\
   \hhline{~~-----------}
   &&BERT &80\% &97\% &87\% &81\% &96\% &88\% &81\% &97\% &89\%&2.87\%\\
   \hhline{~------------}
   &\multirow{2}{*}{\shortstack{Weighted Majority\\(\textit{Labeled Data 4})}}&TF-IDF &78\% &98\% &87\% &80\% &98\% &88\% &81\% &96\% &88\%&4.06\%\\
   \hhline{~~-----------}
   &&BERT &78\% &97\% &87\% &80\% &95\% &87\% &81\% &96\% &88\%&3.61\%\\
   \hline
   \bottomrule
  \end{tabular}
\end{table*}

\begin{figure*}[h]
\centering
\subfloat[]{
\includegraphics[width=0.35\textwidth, keepaspectratio]{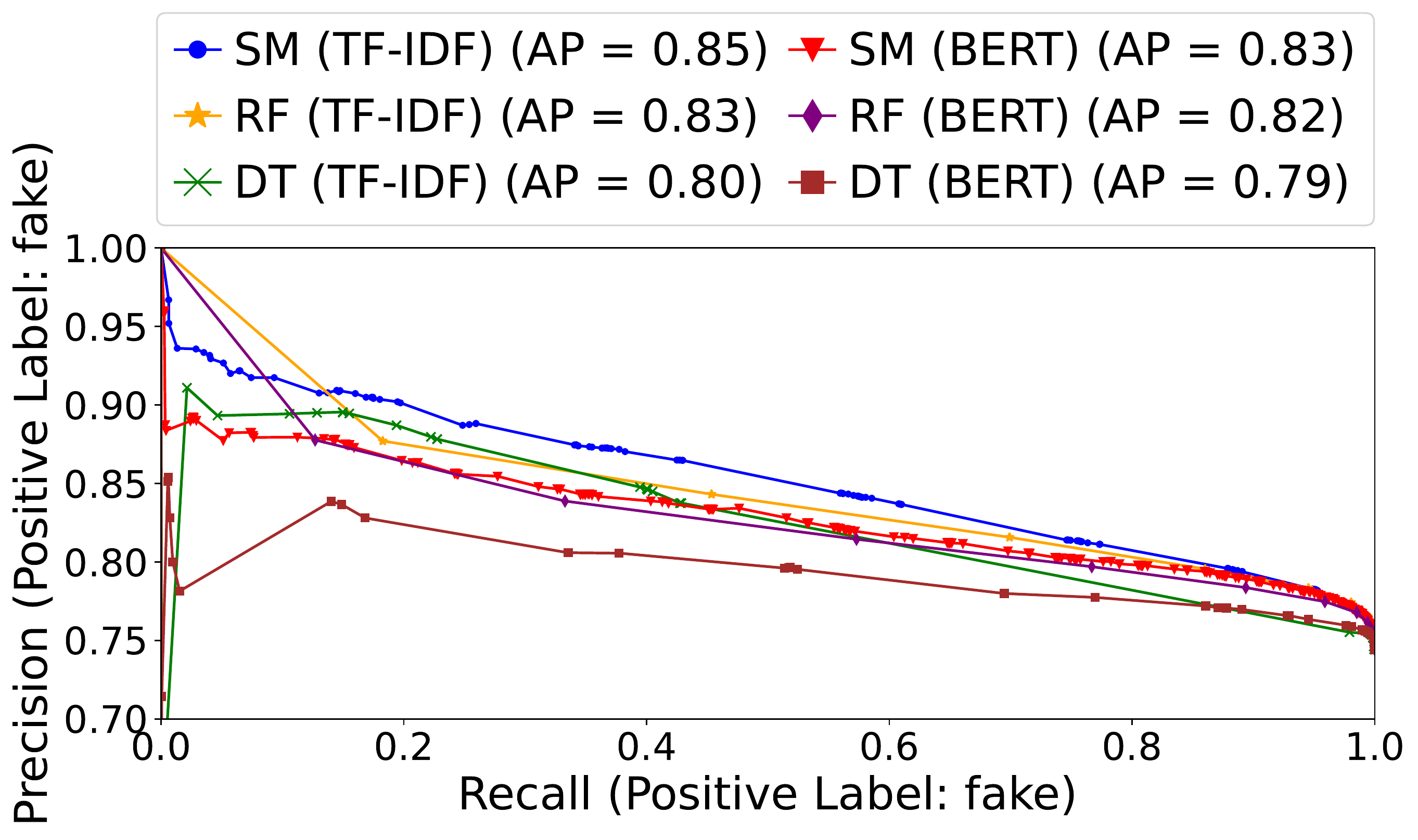} \label{fig:case1a}}
\subfloat[]{\includegraphics[width=0.35\textwidth, keepaspectratio]{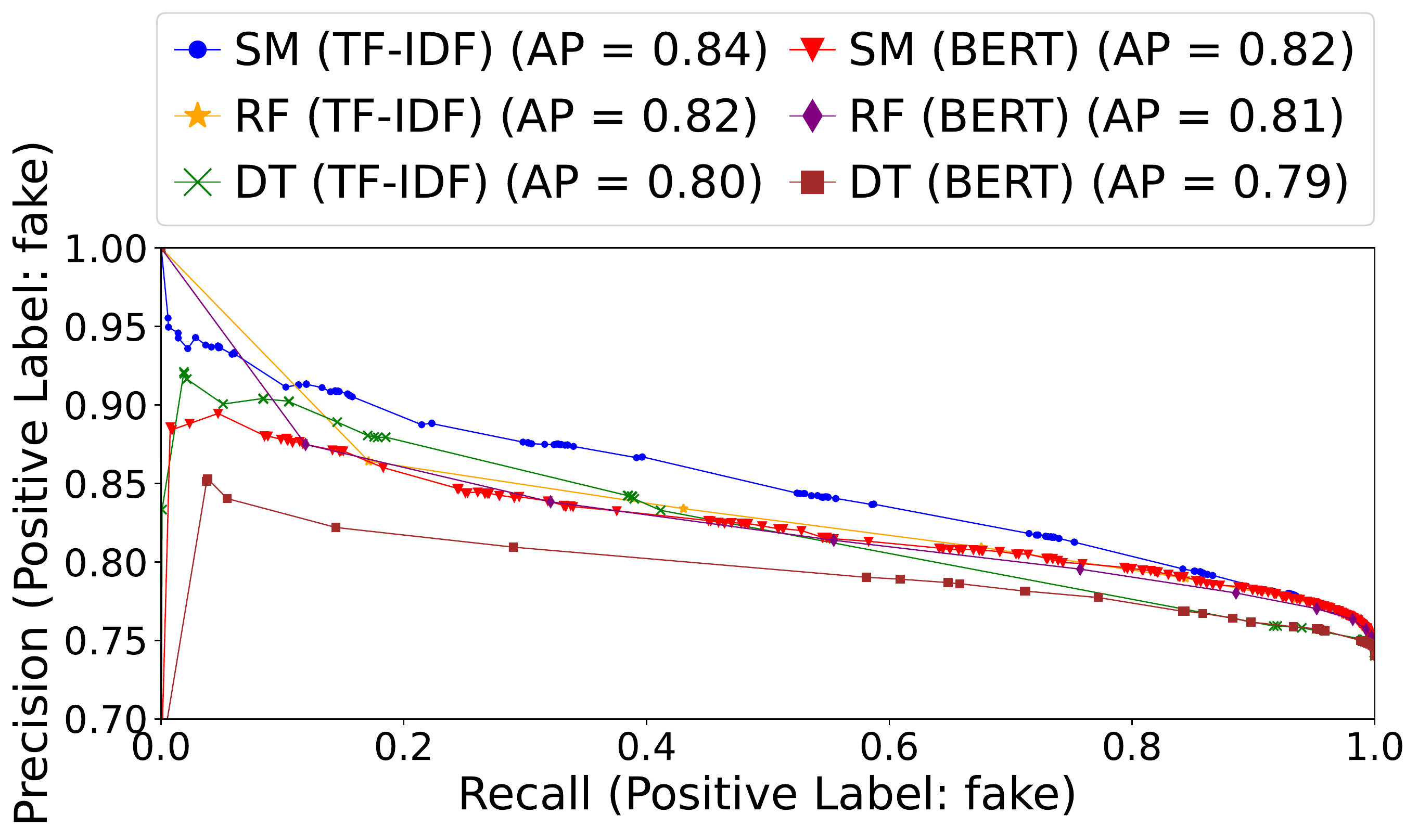} \label{fig:case1b}}\\
\subfloat[]{
\includegraphics[width=0.35\textwidth, keepaspectratio]{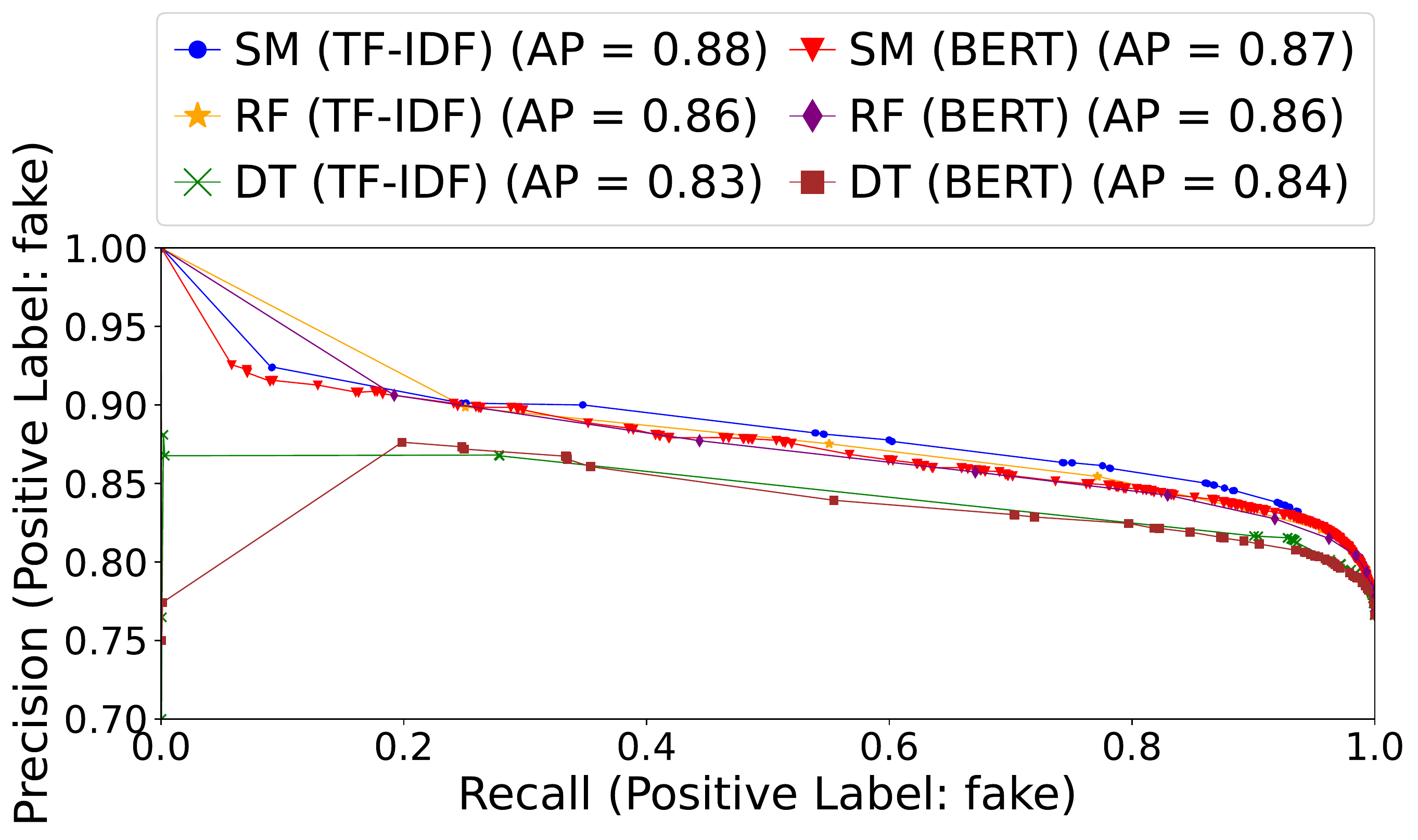} \label{fig:case2a}}
\subfloat[]{\includegraphics[width=0.35\textwidth, keepaspectratio]{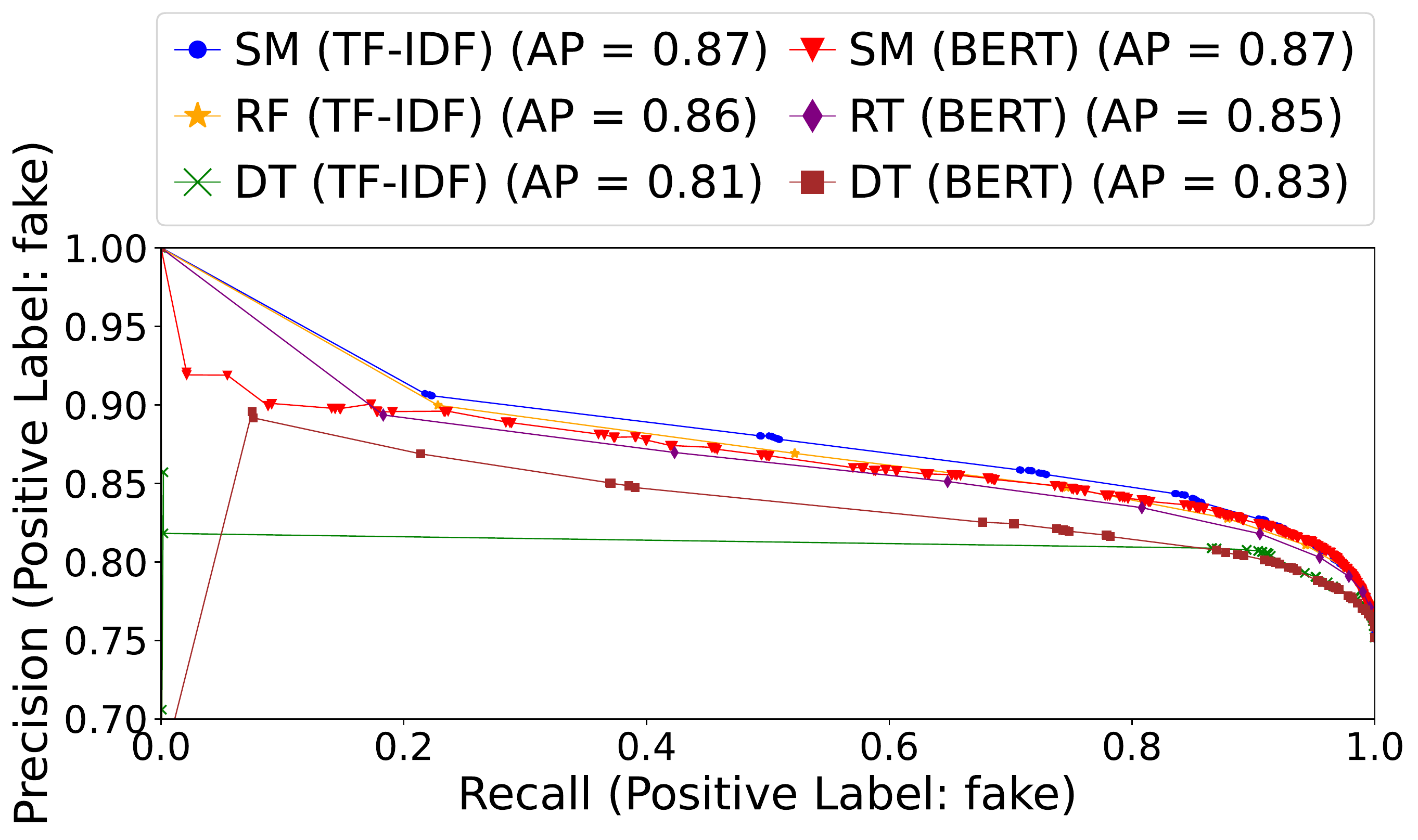} \label{fig:case2b}}
\caption{\small Precision-Recall curve for performance comparison of models on (a) \textit{labeled data 1}  (b) \textit{labeled data 2} (c) \textit{labeled data 3}  (d) \textit{labeled data 4}. Here \textit{SM}, \textit{RF}, \textit{DT} and \textit{AP} represent Stacked Model, Random Forest, Decision Tree and Average Precision.}
\label{fig:case1}
\end{figure*}

At last, our second method of data collection - \textit{with body text} consisting of the supporting statement's title, claim, body content, and verdict helped in increasing the precision and F1-score compared to the datasets \textit{without body text}. As adding more information for the supporting statement adds more context for automated fact-checking. However, \textit{without body text} data was faster to train and test than \textit{with body text} due to having less data for generating the word embeddings. Since our dataset is imbalanced, we have shown the precision-recall curve in Figure \ref{fig:case1}. Figure \ref{fig:case1} compares the performance of the stacked model against the individual classifier with different vectorizers (TF-IDF and BERT). The results indicate that our stacked ensemble model with TF-IDF performs best in all cases in terms of Average Precision (an alternate for the area under curve in the precision-recall curve). In the event-specific dataset, TF-IDF is an efficient text vectorizer and fast.

\section{Bot Detection from Misinformation Tweets}
\label{sec:botdetection}
Unvalidated information on social media portents the dissemination of false information. However, social bots can act as a catalyst in the spread of misinformation \cite{10}. Such bots are automated profiles on OSNs that mimic humans and are easy to generate using the platform's APIs \cite{twitterAPI} and other software. A single botmaster can own multiple bot accounts, acting as a bot army to achieve a malicious goal. For instance, during the 2016 US Presidential Elections, 50,258 Russian bots were found by Twitter that spread disinformation \cite{uselection}. Likewise, in this work, we detect and analyze social bots in the COVID-19 related misinformation tweets. We hypothesize that \textbf{(h1)} bots have an active role in spreading misinformation and \textbf{(h2)} are most active during misinformation campaigns (during peak crisis) compared to other times to maximize their goals. 

Many studies \cite{botometer1,botometer2,botometer3,botometer4} on bot detection have used Botometer \cite{botometer} (a state-of-the-art tool developed at Indiana University) to detect bots. Botometer takes over 1000 features (content, network, sentiment, user features) to calculate a bot score probability between $0$ and $1$. However, Botometer's bot score computation is time-consuming and not scalable for extensive datasets. Due to Twitter API rate limits, we use Botometer Lite \cite{botometerlite}. Botometer Lite is a variation of Botometer that takes in fewer features than Botometer while correlating strongly with Botometer bot scores \cite{thesis}. BotometerLite relies only on features extracted from user metadata contained in the tweet's JSON object when it was extracted. Since user data is already present, it derives other features like tweet frequency, followers count, friends growth rate, etc., to give a bot score without the additional need to query Twitter. This makes it possible to conduct bot detection on collected data. Also, it is scalable to detect bots over large datasets such as our misinformation-labeled dataset.

\begin{figure}[h]
\centering
\subfloat[]{\includegraphics[width=0.23\textwidth, keepaspectratio]{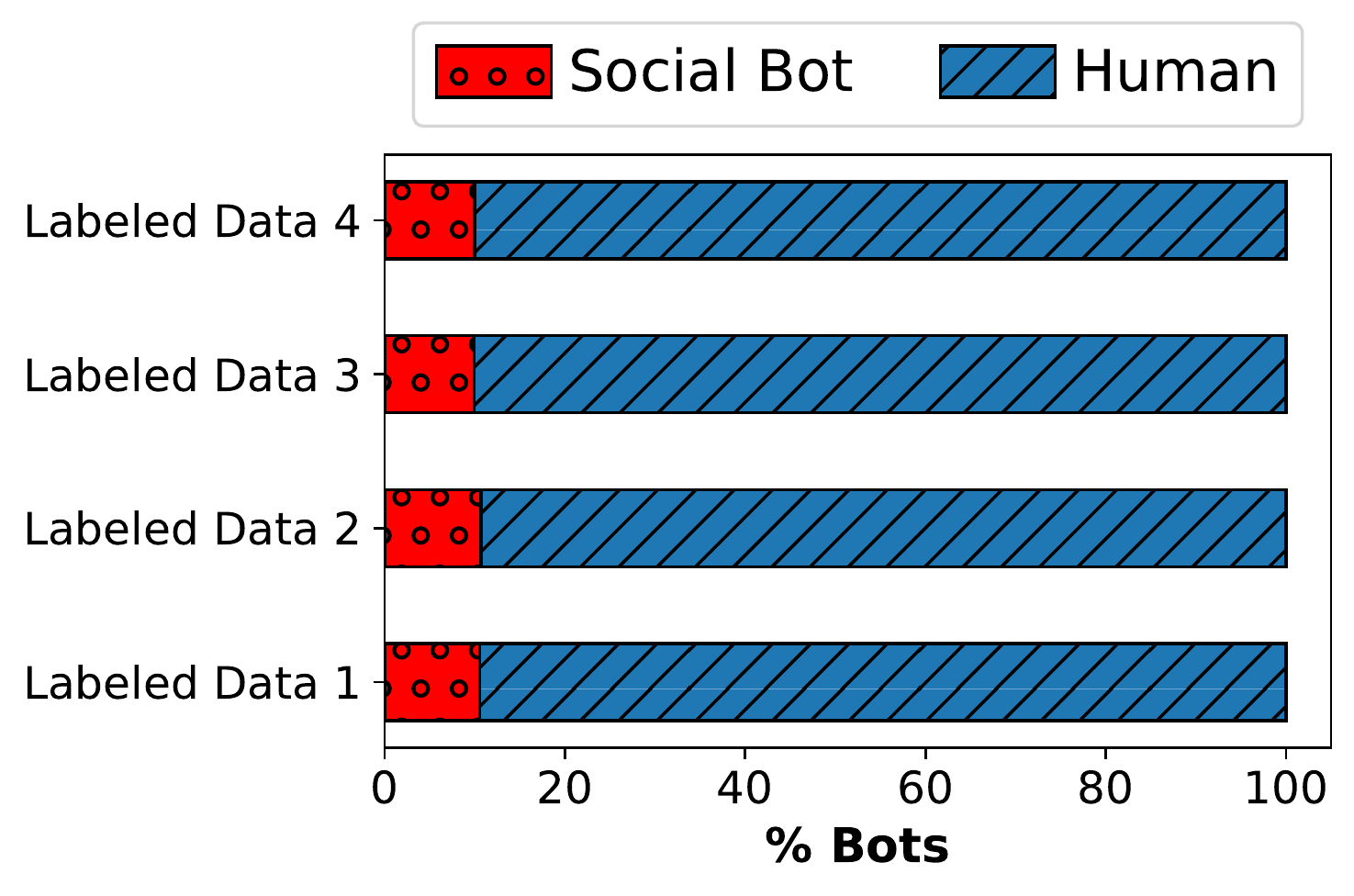} \label{fig:misinfobots}}
\subfloat[]{
\includegraphics[width=0.23\textwidth, keepaspectratio]{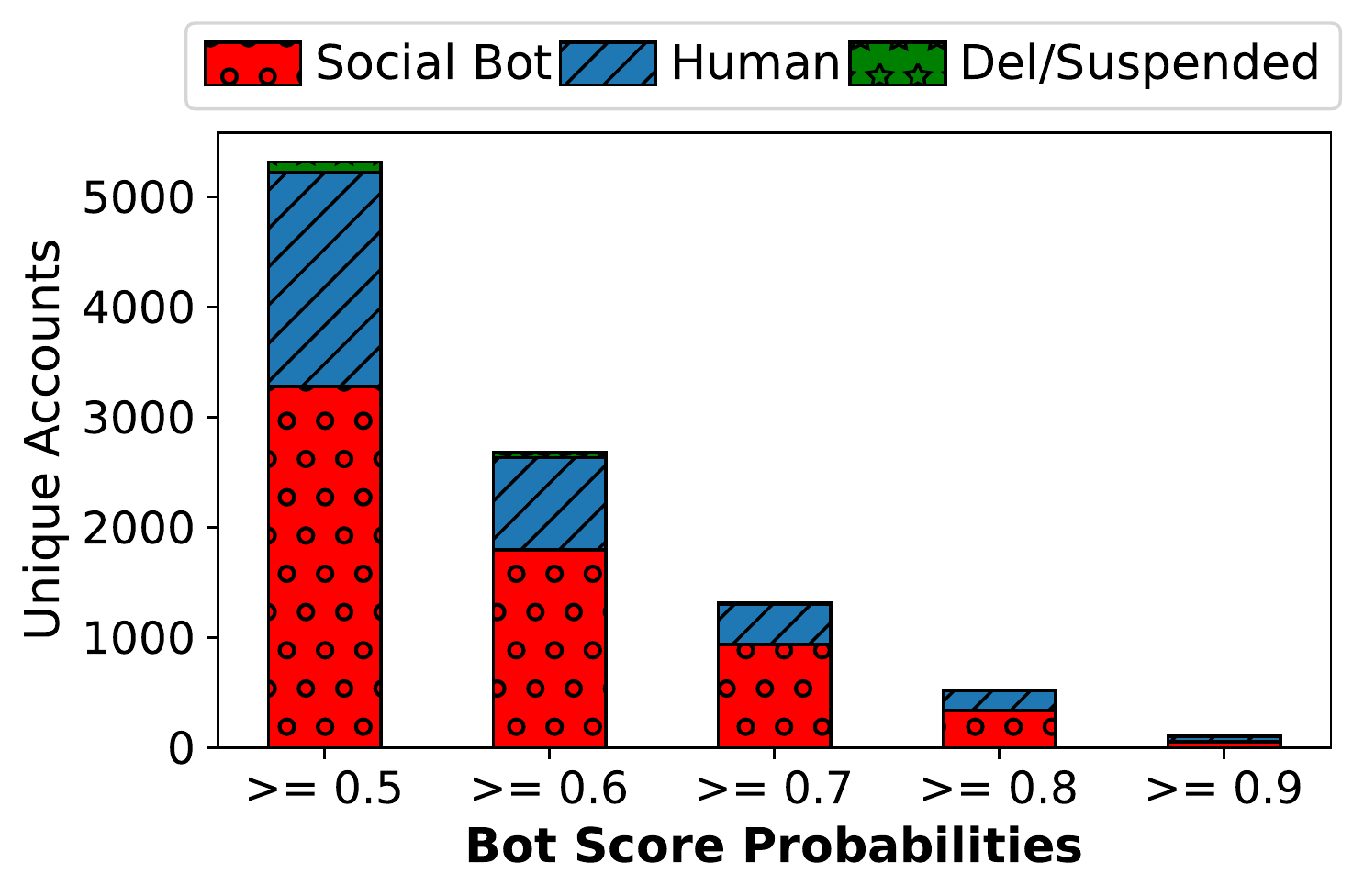} \label{fig:bots_now}}

\caption{\small (a) Presence of {\it bots} in misinformation tweets \textit{(in all four labeled data)} (b) Unique account type in tweets (as of \textit{$13^{th}$ August 2022}).}
\label{fig:bots}
\end{figure}

To begin with, we first need to set the bot score threshold value. If the score is close to $0$, it indicates a human account, and close to 1 means an automated or bot account. In bot detection works, some researchers opt for higher values like 0.7 and 0.8, whereas 0.5 is the more common choice \cite{yang2022botometer}. For example, Shao et al. \cite{threshold1} chose $0.5$ as the threshold between bot and human. Wojcik et al. \cite{threshold2} chose $0.43$, while, in contrast, $0.76$ was deemed as the threshold by Keller and Klinger \cite{threshold3}. Therefore, we have set the threshold as $0.5$. Table \ref{tab:misinformationbots} and Figure \ref{fig:misinfobots} shows the \% of bots in the four labeled datasets with $0.5$ as the threshold. In all four labeled datasets, bots generated approximately 10\% of tweets as shown in Table \ref{tab:misinformationbots} providing evidence to confirm our hypothesis \textbf{(h1)} that \textbf{\textit{bots play an active role in spreading misinformation in online social networks}}.  

 \begin{table}[!ht]
\centering
\caption{Number of {\it bots'} generated tweets in four labeled, COVID-19 related misinformation, datasets.}
\scalebox{1.0} {\begin{tabular}{c|c|c|c}
\toprule
\hline
\textbf{Data} & \textbf{\# Fake Tweets} & \textbf{\# Bot-generated Tweets} & \textbf{\%} \\ \hline
Labeled Data 1                      & 93,348              & 9,885   & \textbf{10.58\%}                             \\ 
Labeled Data 2                      & 92,619              & 9,894     & \textbf{10.68\%}                           \\ 
Labeled Data 3                      & 96,461              & 9,574 & \textbf{9.92\%}                                 \\ 
Labeled Data 4                      & 94,600              & 9,417 & \textbf{9.95\%}                                 \\ 
\hline
\bottomrule
\end{tabular}}
\label{tab:misinformationbots}
\end{table}

We used 125,709 tweets that were crawled during June-August 2021 for labeling. Those tweets had been tweeted by 69,665 unique users. As of August 2022, 122,314 tweets out of that are still available (3,395 tweets are either deleted by users or removed by Twitter) and they have been tweeted by 68,377 unique users. Table \ref{tab:unqiuebots} shows the comparison of bot statistics during June-August 2021 and August 2022. With the threshold of 0.5 ($bot score>=0.5$), 5,315 (out of 69,665) accounts were identified as bots during June-August 2021. By August 2022, out of the 5,315 bot accounts, 3,279 were still acting as bots, 1,941 were acting as genuine OSN (human) accounts ($bot score<0.5$), and 95 user accounts have been deleted or suspended. Figure \ref{fig:bots_now} illustrates the number of accounts showing bot behaviour and genuine OSN user behaviour in August 2022 for various bot scores (compared to the number of bot accounts identified in June-August 2021). It is evident from Table \ref{tab:unqiuebots} that now one-third of bot accounts display genuine user (human) activity in the August 2022 data. We can use this result to show that bots have maximum motive to spread misinformation during a crisis period like the peak of COVID-19 pandemic compared to normal times. This change of bot's behavior affirms our hypothesis \textbf{(h2)} that \textbf{\textit{bots are most active during their misinformation campaigns compared to other times}}. As confirmed from our hypotheses (h1) and (h2), bots have played a role in promoting misinformation during COVID-19.

\begin{table*}[h]
  \centering
  \caption{Number of {\it unique bot accounts} in misinformation data across different {\it bot score} probability threshold values. }
\label{tab:unqiuebots}
  \begin{tabular}{c|c|c|c|c}
  \toprule
  \hline
  \multirow{2}{*}{\textbf{Thresholds}}&\multirow{2}{*}{\shortstack{\textbf{\# Unique Accounts (as of June-August 2021)}}}
&\multicolumn{3}{c}{\shortstack{\textbf{\# Unique Accounts (as of August 2022)}
}}\\\hhline{~~---}
  & &\textbf{\# Bot Accounts} & \textbf{\# Human Accounts} &\textbf{\# Deleted/Suspended Accounts}\\
  \hline
  \textbf{$>= 0.5$}&\textbf{~~~5,315}& \textbf{3,279 (61.7\%)} & \textbf{1,941 (36.5\%)} & \textbf{95 (1.8\%)} \\
  $>= 0.6$& ~~~2,680 & 1,796 (67.0\%) & ~~838 (31.3\%) & 46 (1.7\%) \\
  $>= 0.7$& ~~~1,316 & ~~936 (71.1\%) & ~~361 (27.4\%) & 19 (1.4\%)\\
  $>= 0.8$& ~~~~~526 & ~~336 (63.9\%) & ~~184 (35.0\%) & ~6 (1.1\%)\\
$>= 0.9$& ~~~~~103 & ~~~~54 (52.4\%) & ~~~~48 (46.6\%) & ~1 (1.0\%)\\
  \hline
  \hline
  \textbf{\# Tweet Sample}&125,709&\multicolumn{3}{c}{122,314}\\
  \hline
   \textbf{\# Unique Users}&~69,665&\multicolumn{3}{c}{~~68,377}\\
  \hline
  \bottomrule
  \end{tabular}
\end{table*}

\section{Related Work}
\label{sec:relatedwork}

Since the beginning of the pandemic, several works have analysed myths circulating on COVID-19, specifically on social media. Yang et.al. \cite{yang2020analysis} analysed myths circulating on Twitter by classifying them into five different categories, followed by fitting them with the Susceptible-Infectious-Recovered (SIR) epidemic model which characterizes the basic reproduction number $R_0$ for each category. $R_0 > 1$, indicates high probability of an infodemic. The study also analysed the public's emotions associated with each category of myths. While the study found that the misinformation regarding the spread of infection and preventive measures communicated much faster than  other categories ($R_0 > 1$ for these categories), it also revealed fear being the dominant emotion in all categories with 64\% of the collected tweets expressing the emotion of fear. 

Cinelli et al. \cite{cinelli2020covid} measured people's interaction and engagement with COVID-19 infodemics by modeling the spreading information with epidemic models distinguishing the basic reproduction number $R_0$. For research, approximately 8 million posts from five different social media platforms, including, Gab, Reddit, Instagram, Youtube, and Twitter were collected between 1 January 2020 and 14 February 2020. The methodology of the research includes calculating the cumulative number of posts and the number of reactions to each post, followed by estimation of the basic reproduction number $R_0$ for each platform using the phenomenological model of \cite{fisman2013idea} and the compartmental model (Susceptible,
Infected, Recovered). The study revealed that the values of $R_0$ are significantly greater than 1, which indicates the prevalence of infodemics on social media platforms. Based on the methods and results of \cite{yang2020analysis, cinelli2020covid}, their techniques can be considered reliable enough to analyze myths circulating about COVID-19. 

Shu et al. \cite{shu2017fake} explored challenges associated with misinformation detection, existing algorithms to identify them from a data mining perspective, and the relevant datasets. The survey characterises misinformation based on psychology foundations, social foundations, malicious accounts, and echo chambers. The research also provides different detection methods, including, knowledge-based detection, style-based detection, stance-based detection, and propagation-based detection. However, our work is close to the post-based (social context) method.

Alonso et al. \cite{DBLP:journals/corr/abs-2112-04831} showed that BERT is the best model for text feature extraction for unimodal data, and CNN works well for multimodal data. They, however worked on multiclassification rather than binary classification. In their work, five classes were used: true, manipulated content, false connection, satire, and imposter content. However, their dataset (Fakeddit dataset) is an imbalanced dataset with majority of data falling in the category of true news which does not help in understanding characteristics of imposter content or false news. 

To predict future fake news topics, Vicario et al. \cite{vicario2019polarization} proposed a framework to identify polarizing content on social media. They used the Italian Facebook dataset to validate the performance of their proposed framework. The framework consists of four main phases: data collection, topic extraction and sentiment analysis, features definition, and classification. Upon validating the proposed methodology, it was found that the framework could identify topics vulnerable to misinformation with a 77\% accuracy. However, the dataset is moderate size but contains fewer fake samples (only 17 Facebook pages for the fake category compared to 58 pages for the real category). Also, this research has another limitation because they assume that all false news originates from unofficial news sources, which is not always the case since unsubstantiated information diffuses even from official newspapers. 

Similarly, Uddipta et al. [24] leveraged multiple supervised classifiers to detect anti-vaccination tweets using only 75 manually annotated anti-vaccination tweets. Most of the previous work was limited in performance due to the small annotated datasets. Manual data annotation is an expensive task as it requires expertise, time, and cost. The annotation process is not scalable concerning the number of fact-checkers available. This alone the limits the work of fake news detection. Since more available data means more patterns of misinformation can be understood. In the absence of an extensive data set, most of the indicators are lost, and researchers have to rely on patterns from a small dataset, leading to a poor predictive model. Therefore, our end-to-end annotation and labeling method helps build an extensive dataset for fake news detection research.

On the other hand, other work primarily focuses on fake news detection in ``news articles" rather than social media posts \cite{Liar}. For instance, Wang \cite{Liar} used hybrid-CNN to detect fake news in short political statements extracted from Politifact. Similarly, Aldwairi and Alwahedi \cite{ALDWAIRI2018215} used syntactical structure of the website links and words of the news titles to detect clickbaits and fake news. Shu et al. \cite{fakenewsnet} focuses on identifying the partisanship of news publishers, and Wang et al. \cite{wang2020weak} used reinforcement learning
to detect fake news in new articles based on user reports. Our work rests differently than others as we solely identify misinformation in social media posts instead of fake news articles or URL that link to external websites. Our work focuses on the text of the post that is used for spreading misinformation since social media is conversation-based platform.

\section{Conclusion}
\label{sec:conclusion}
High consumption and the ready availability of news on social media influence people on many major topics. In recent years, social media has exposed OSN users to misinformation, causing the rapid spread of infodemic. While fake news detection is not a new concept, there exists limited work on annotating ground truth data. In this paper, we proposed an \textit{annotation model} for creating a ground truth COVID-19 tweet dataset and an \textit{ensemble stack model} machine learning-based classifier to detect fake news. Our model achieves a \textit{precision} score of 82\% and \textit{recall} score of 96\% when tweet's textual features are extracted using TF-IDF. Both TF-IDF and BERT transformer model performed almost same on all four labeled datasets, we created. Moreover, increasing body context information of supporting statements (\textit{with body text}), achieved any performance improvement compared to supporting statements with only title and verdict (\textit{without body text}). We further outlined evidence that bots play an active role in spreading misinformation by analyzing the bot's presence using the Botometer Lite tool. Bots generated approximately 10\% of misinformation tweets. We also showed that bot's behavior changes over time, depicting that bots are most active during their misinformation campaigns compared to other times. Our methodology and classification model helps researchers annotate datasets and detect fake news, thus increasing users’ trust in information on OSN.

\textit{Future Work:}  
Filtering tweets with a claim, which is worth fact-checking can improve the categorization of the misinformation dataset. Claim identification or stance detection can provide a basis for such extensive filtering. Moreover, our work can leverage entity-based detection or event-invariant features to detect fake news in unseen events as future work. Such a model can classify fake news on new events using an event discriminator to learn shared features and remove event-specific features. Further research can progress on how misinformation and bots fit into more extensive campaigns intending to carry out malicious activities.

\balance\bibliographystyle{IEEEtran}
\bibliography{main}

\end{document}